\documentclass[10pt]{iopart}
\usepackage{psfig}
\begin{document}

\title[Analysis of kaon spectra at SIS energies -
what remains from the KN potential?]{
Analysis of kaon spectra at SIS energies - \\
what remains from the KN potential?  
\footnote[1]{Supported by IN2P3/CNRS and GSI.}}

\author{C Hartnack  
and J Aichelin }

\address{
SUBATECH,  UMR 6457, Ecole des Mines de Nantes, IN2P3/CNRS et Universit\'e de
Nantes,  4, rue A. Kastler, B.P. 20722,
44307 Nantes, France}

\begin{abstract}
We study the reaction Au+Au at 1.48 AGeV and analyze the influence of the
KN optical potential on cm spectra and azimuthal distributions at mid-rapidity.
We find a significant change of the yields but only slight changes in the shapes
of the distributions when turning off the optical potential. 
However, the spectra show contributions from different reaction times, where
early kaons contribute stronger to higher momenta and late kaons to lower momenta.
Azimuthal distributions of the kaons at mid-rapidity show a strong centrality
dependence. Their shape is influenced by the KN optical potential as well as
by re-scattering.  
\end{abstract}

\vspace*{-3mm}

\section{Introduction}
One key question in the analysis of sub-threshold kaon production 
is how to obtain information on the properties of strange mesons in dense nuclear 
matter. Especially the  relation of the optical potential of $K^+$ 
and $K^-$ in nuclear medium  to experimental observables like
$K^-/K^+$ ratios, mesonic in-plane flow and azimuthal distribution of kaons
is subject of vivid discussions, who have triggered a lot of activities on the
experimental \cite{Barth,Menzel,Ritman,Crochet,Devismes,Laue,Sturm,Foerster}
and theoretical side 
\cite{schaffi,Cassing,CLE00,ho_s2000,Ko01,Fuchs,LiKo,Wang}.

In this article we study the production of $K^+$ in the reaction Au+Au at 
1.48 AGeV which corresponds to recent experiments performed by 
the KaoS collaboration \cite{Foerster} and by the FOPI collaboration \cite{Devismes}.
For this purpose we use the IQMD model \cite{iqmd,bass}
where we have supplemented our standard simulation program by  all 
relevant cross sections for kaon production and annihilation and a 
(density and momentum dependent) $KN$ optical potential.  
For the latter we use a parametrization resulting from 
relativistic mean field calculations of Schaffner-Bielich \cite{juergen}. 
Detailed description of our simulations can be found in \cite{ikaon,kminus}.

\section{Time evolution of kaon production}
\begin{figure}[hbt]
 \centerline{\psfig{file=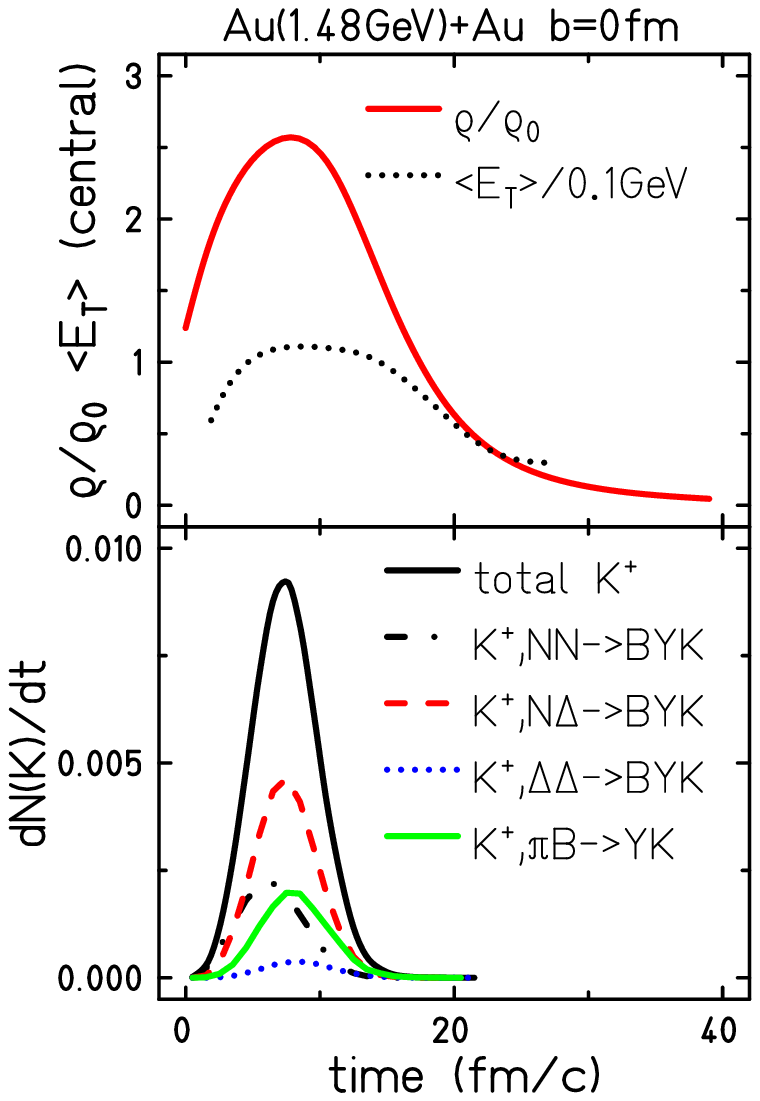,width=0.4\textwidth,angle=-0} \
 \psfig{file=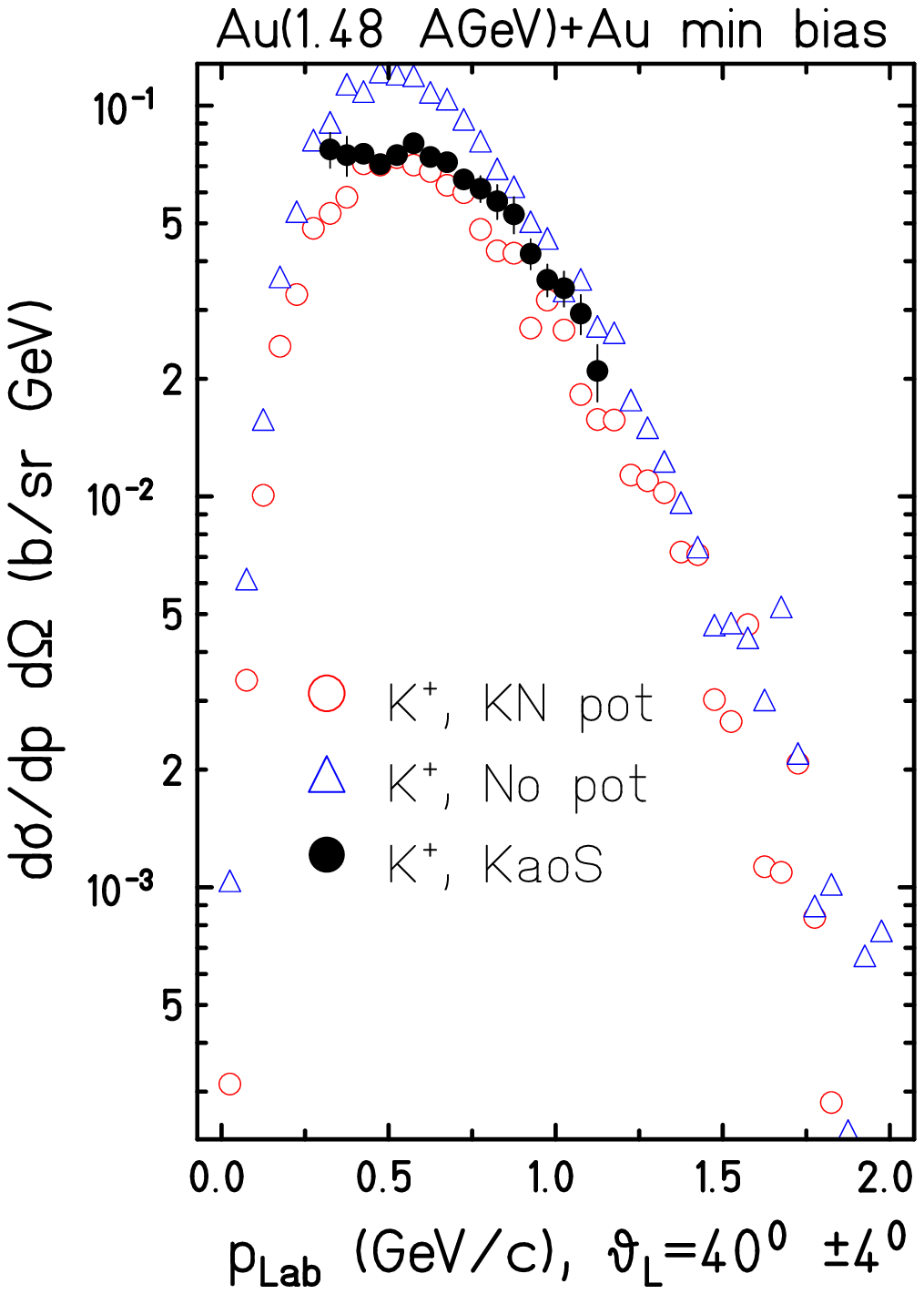,width=0.42\textwidth,angle=-0} }
\caption[Time evolution of density and kaon production and lab spectrum at
 $40^\circ$]{Left: Time evolution
of the density (upper part) and of the production of kaons (bottom) and
the contribution of different channels \\
Right: Lab momentum spectra of a minimum bias Au(1.48 AGeV)+Au  
collision at $40^\circ$ with and without KN potentials compared to 
KaoS Data \cite{Foerster}.
}
\label{timeevol}
\end{figure}

Let us first take a short view on the reaction dynamics of a central
Au+Au collision at 1.48 AGeV. Figure \ref{timeevol} shows on top of the left hand side
the time evolution of the normalized baryonic density (full line) 
taken at the center of the  reaction in a  b=0 fm collision. 
We see a maximum at about 8 fm/c after first contact (0 fm/c) 
yielding about 2.5 times ground state densities.
In the same figure the dotted line shows the time evolution of the transverse
energy at the center of the reaction which can  be taken as an indicator of 
temperature. The transverse energies are divided by 0.1 GeV, thus 
a value of 1 corresponds to 100 MeV.
We see a strong rise at very early times (up to about 4 fm/c) followed by a 
plateau and a moderate decrease.  
We see that $K^+$ are produced quite early (dashed line) and that the maximum of their
production co-incidents with the maximum of baryonic densities. 

The bottom part of the left hand side of Figure \ref{timeevol} shows the time 
evolution of the $K^+$ production and its decomposition into different channels.
the different production channels. 
We see that $K^+$ are produced quite early (full black line
) and that the maximum of their
production co-incidents with the maximum of baryonic densities. 
The different channels show a rather similar shape.
The production of $K^+$ in direct $NN$ collisions (dash-dotted line) 
acts earliest. It shows it maximum at about 4-6 fm/c, a time when the
first violent reactions already took part and the transverse energy (left
hand side, dotted line) reached its plateau. At that time first equilibration
effects in the collision dynamics can be assumed. The production of $K^+$ in
$N\Delta$ (dashed line) and $\Delta\Delta$ collisions shows it maximum a little
bit later. These collisions may dominantly take place at high densities in the
stopped matter when some equilibrium is already reached. The $\pi B$ channel
(full gr. line) finally peaks at still later time due to the delayed freeze-out
of pions.

\section{Analysis of kaon spectra}

If the kaons are produced at highest densities we may assume that they 
should carry a signature of the KN potentials. Indeed the penalty of the
KN-potentials can clearly be seen in the absolute yield of the produced
$K^+$\cite{Cassing,Ko01,Fuchs,ikaon,sqm2001}.
This penalty shifts the threshold for the
production toward higher values and therefore reduces the production yields.
That effect can be easily seen on the right hand side of figure \ref{timeevol}
where we compare laboratory momentum spectra taken at $40^\circ$ laboratory angle
taken from calculations with (open circles) and without
(open triangles) KN optical potential to recent data
taken by the KaoS collaboration (bullets) \cite{Foerster}.
For our calculations we assumed a total cross section of 5.3 barn for minimum 
bias reactions.
We see that a calculation without KN potentials yields higher yields than
a calculation with potential. The data favor the calculation with
KN potentials. Nevertheless it should be noted that incertitudes on unknown
production cross sections like $N\Delta\to NYK$ may alter this statement
\cite{sqm2001}. 

\begin{figure}[hbt]
 \centerline{\psfig{file=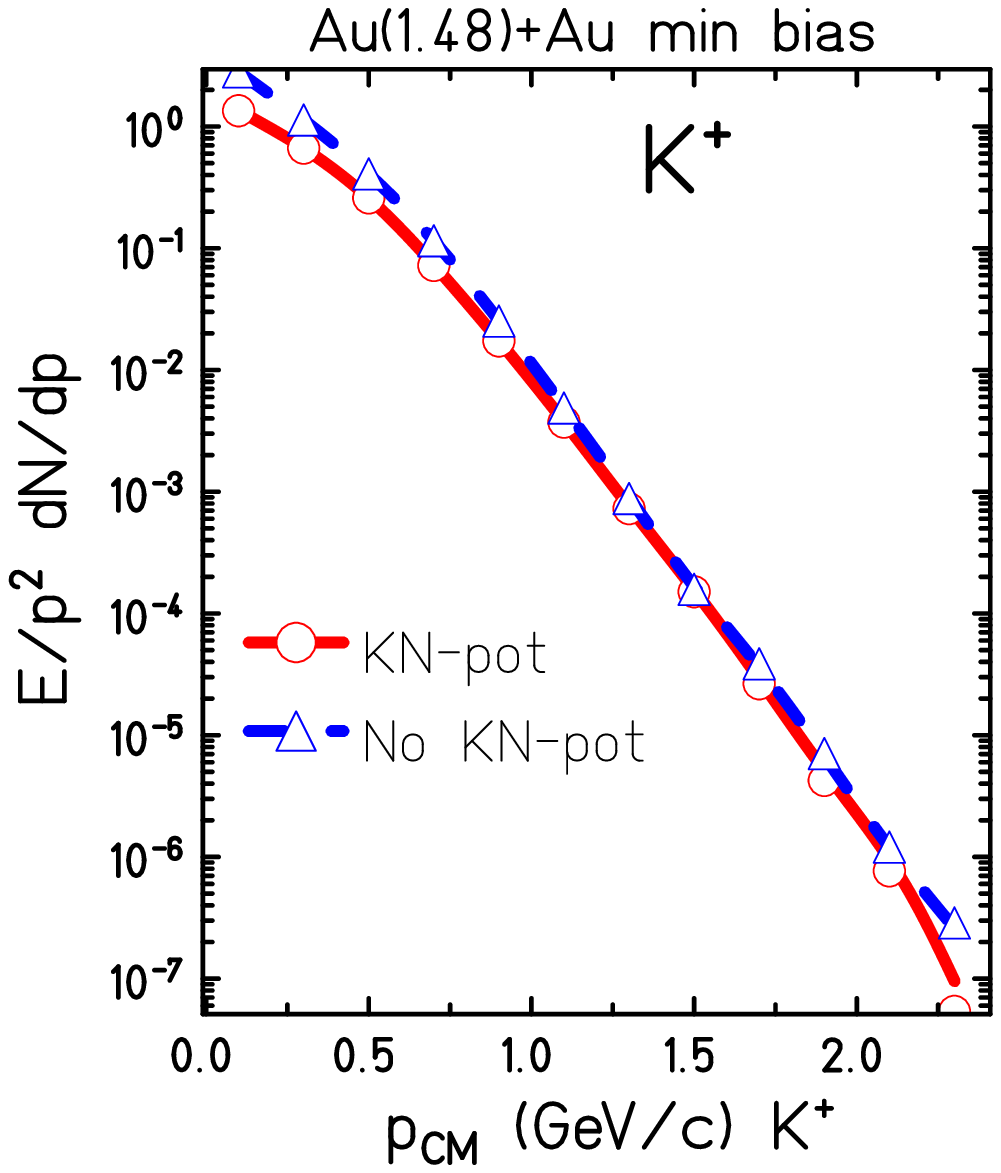,width=0.4\textwidth,angle=-0} \
 \psfig{file=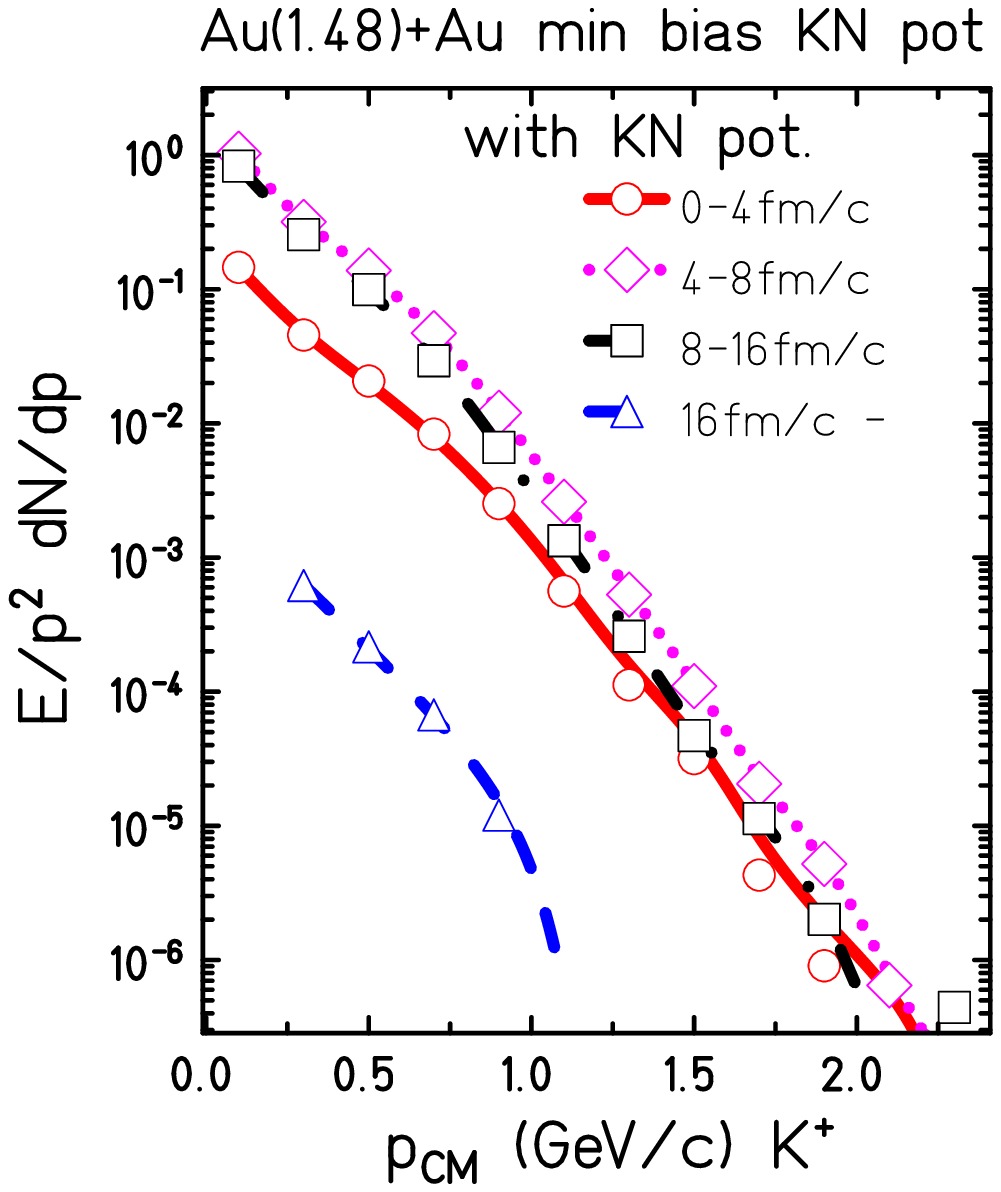,width=0.4\textwidth,angle=-0} }
\caption[Kaon spectra with and without potentials and time decomposition of 
kaon spectra]{Left: Spectra of kaons in minimum bias reactions calculated with 
and without KN potential \\
Right: Decomposition of kaon spectra into contributions of different 
reaction times.
}
\label{spectra}
\end{figure}

In order to allow the analysis of high energetic kaons let us now turn to
center-of-mass momentum spectra in full $4\pi$ as they are shown in figure 
\ref{spectra}. As we already stated, the penalty of the KN potentials 
shifts the threshold for the production toward higher values.
Furthermore the produced kaons get less energy from the producing collision.
However, this energy is regained in the expansion when the kaon leaves the
nuclear medium. Therefore it is not very surprising to see on the left hand 
side of figure \ref{spectra} that the spectrum of the $K^+$ in a calculation with 
potential (full line with circles) does not very much differ to that in a
calculation without KN-potentials (dashed line with triangles), especially at
high kaon momenta. Remember that the energy range shown in fig. \ref{spectra}
is larger than that shown on the right hand side of fig. \ref{timeevol}.
Only at low momenta we see a difference in the absolute values. This 
difference corresponds to kaons which are produced slightly above the threshold.
In calculations without potentials they can be produced but they get little 
momentum for kinematic reasons. In calculations with potential their production
is forbidden due to the enhanced threshold.
For high kaon momenta re-scattering plays an important role. If we disable kaon
re-scattering the values at high momenta drop down visibly.

The right hand side of figure \ref{spectra} decomposes the spectrum obtained
in a calculation with KN potentials into different time windows. 
The time-window $t=$0-4 fm/c (full line with circles) corresponds to the violent
non-equilibrated phase where density and transverse energy is rising (a 
temperature cannot be defined at that time). We see a strong contribution
at high cm momenta. When we go to later time windows the spectra are becoming
steeper. The window $t=4-8$fm/c (dotted line with diamonds) corresponds to the
time when equilibrium is starting but compression is still increasing. It shows
the dominant contribution to the overall spectrum. In the expansion state
(t=8-16fm/c, dash-dotted line with square) a strong contribution can be found 
at lower momenta. The very late kaons ($t>16$ fm/c, dashed line with triangles)
finally do not contribute significantly.
This effect is coherent with the observation that re-scattering is important for
the high momentum part of the spectrum. The later the kaons are produced, the
less time and less density they find for doing re-scattering.
The behavior ``the later the
production the steeper the spectra'' is already known from the analysis of
pion spectra \cite{bass} and from perturbative studies of kaon production 
\cite{oldhartnack}. It indicates that the detailed analysis of the spectra 
gives information on the time structure of the collision dynamics rather than 
on the KN potentials.

\section{Azimuthal distributions at mid-rapidity}

\begin{figure}[hbt]
 \centerline{\psfig{file=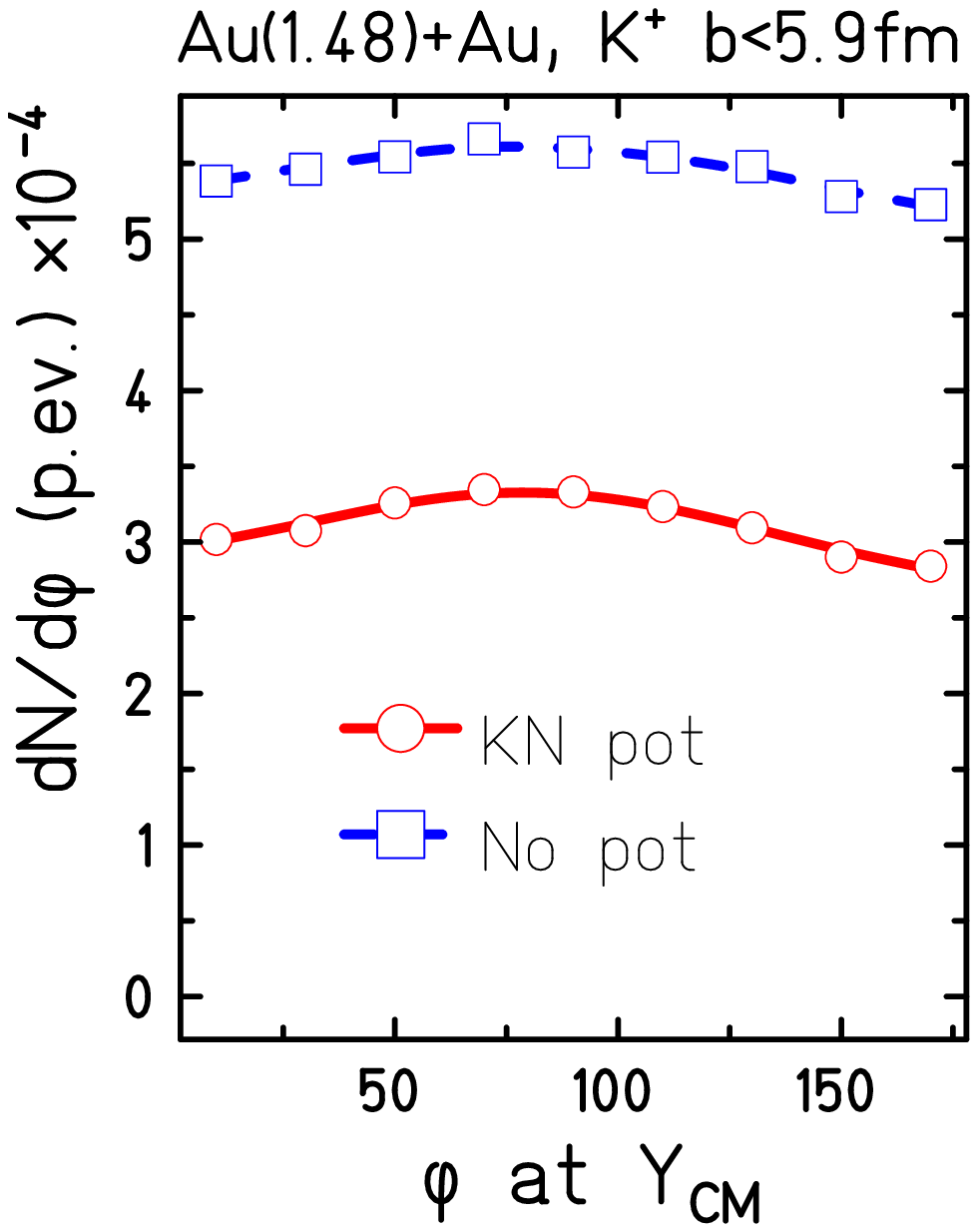,width=0.4\textwidth,angle=-0} \
 \psfig{file=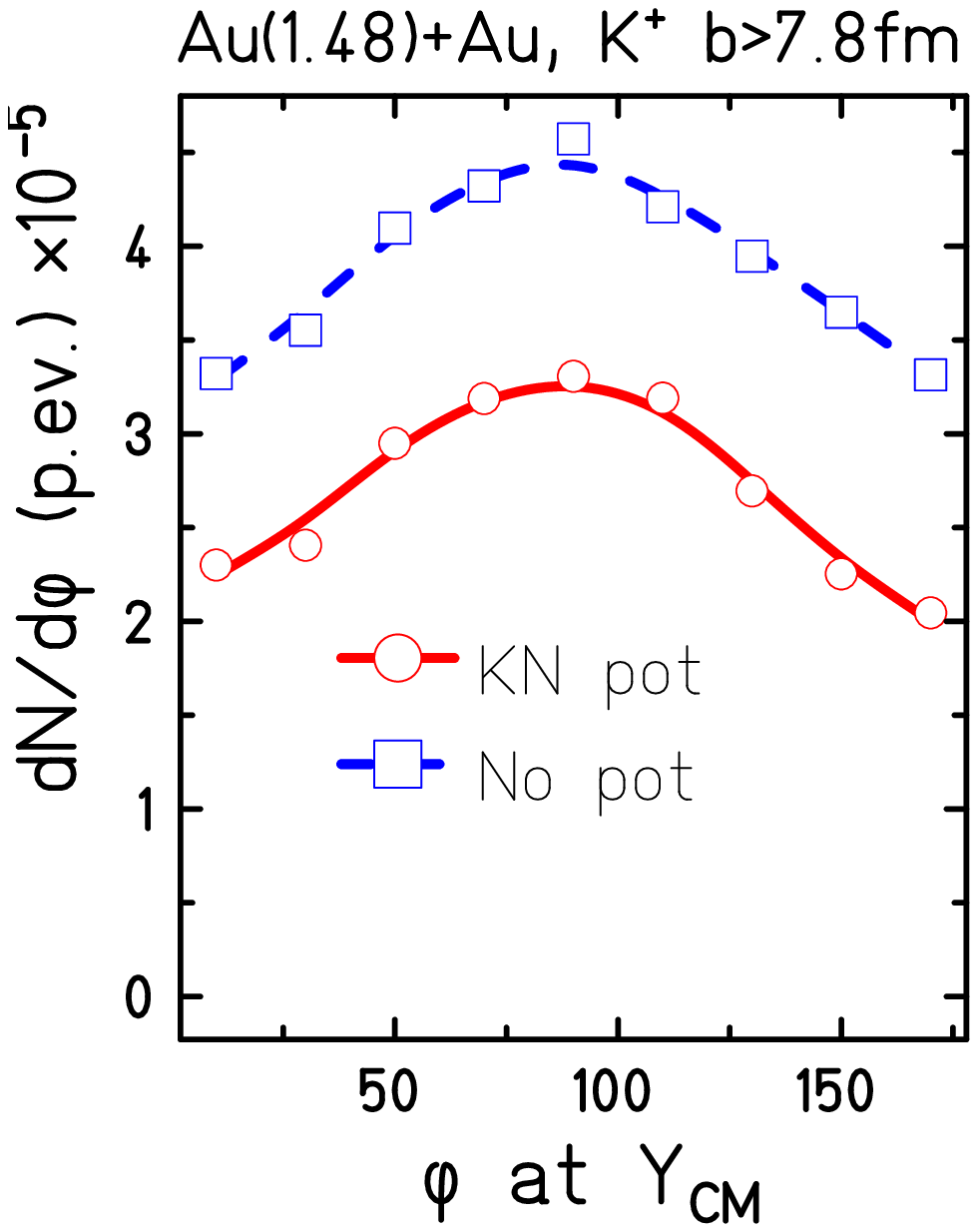,width=0.4\textwidth,angle=-0} }
\caption[Azimuthal distributions of kaons at mid-rapidity in central and peripheral 
collisions]{Azimuthal distributions of kaons at mid-rapidity calculated with
and without potential in central (left) and peripheral (right) collisions
}
\label{squeeze-b}
\end{figure}
Let us now turn to the azimuthal distributions of kaons at mid-rapidity.
For this observable an influence of medium effects has been reported by
\cite{LiKo,Wang}. Fig \ref{squeeze-b} shows the azimuthal distributions 
for calculations with KN potentials (full line with circles) and without
KN potentials (dashed line with squares) for central ($b<5.9$fm, left hand side)
and peripheral ($b>7.8$fm, right hand side) collisions. We see a strong 
dependence of the centrality. While central collisions only show a slight 
maximum at $90^\circ$, i.e. out of the reaction plane, this maximum becomes
a significant peak for peripheral collisions. This centrality dependence
indicates the importance of the spectator matter for understanding this
behavior. We have to discuss whether it is the potential or the re-scattering
in matter which is important.

\begin{figure}[hbt]
 \centerline{\psfig{file=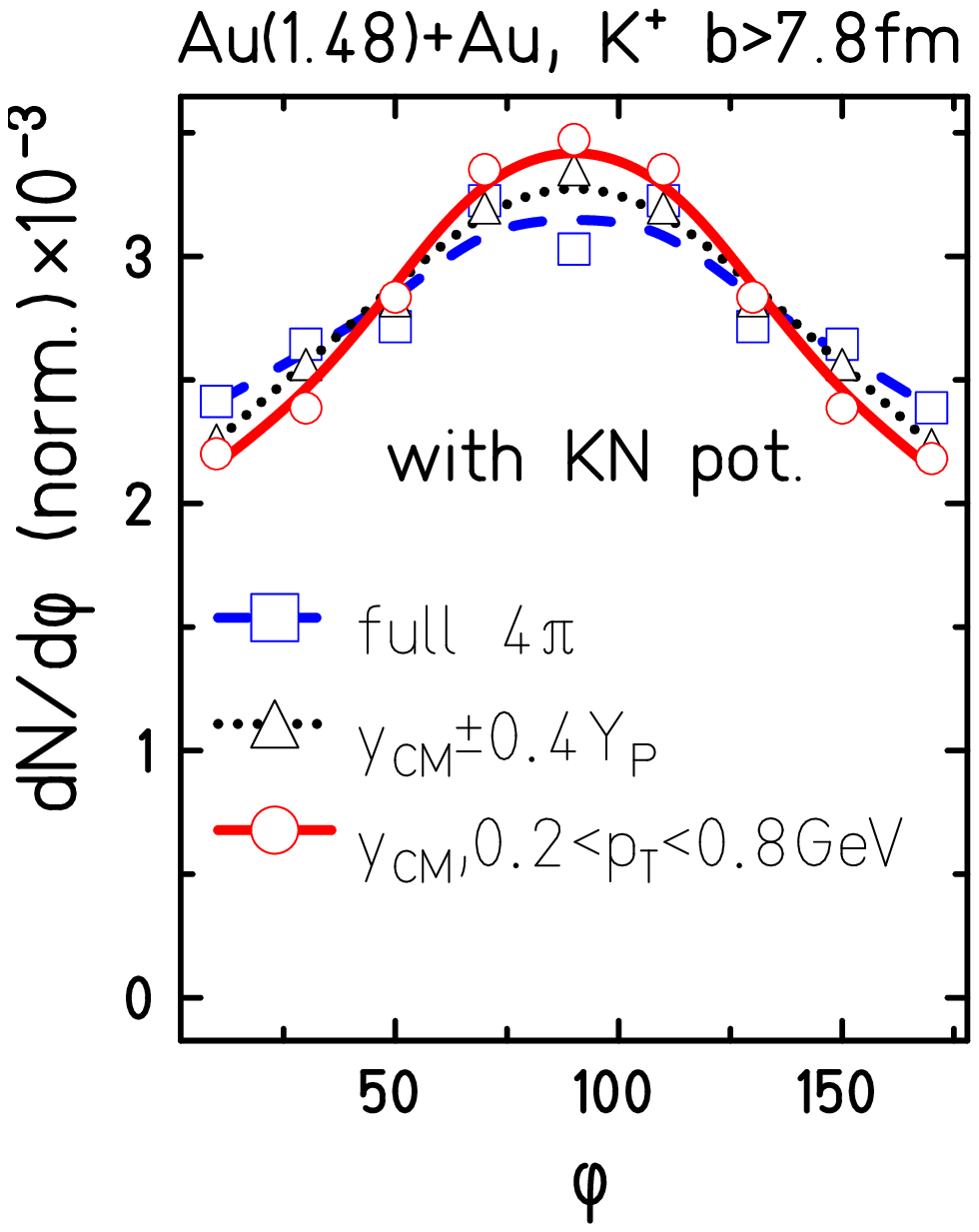,width=0.4\textwidth,angle=-0} \
 \psfig{file=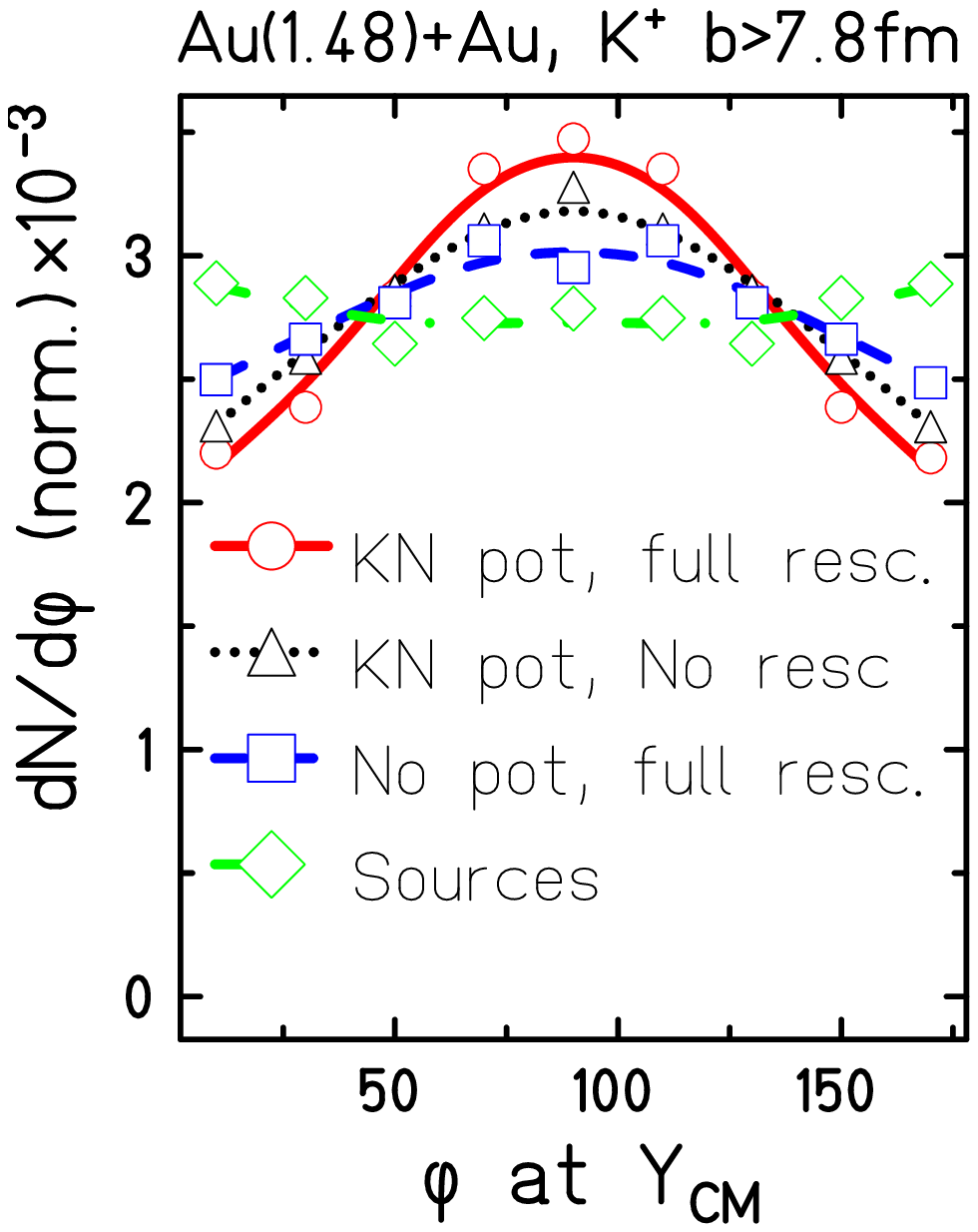,width=0.4\textwidth,angle=-0} }
\caption[Normalized azimuthal distributions of kaons at mid-rapidity in peripheral 
collisions]{
Left: Normalized azimuthal distributions for all rapidities, at 
mid-rapidity and at mid-rapidity with additional $p_T$ cuts. \\
Right: Normalized azimuthal distributions of kaons at mid-rapidity 
calculated in peripheral collisions  with 
and without potential and with and without re-scattering  
}
\label{squeeze-p}
\end{figure}
Concerning the potential we already see a large influence on the absolute
yield of the distribution. In order to compare the shapes of the curves 
figure \ref{squeeze-p} shows normalized distributions for peripheral collisions.
In central collisions the maximum of the distribution is less significant.
 
The left hand side of figure \ref{squeeze-p} shows the influence of different
cuts which have been used by experiment\cite{Foerster}. The dashed line with squares shows the
distribution integrated over all rapidities, the dotted line with triangles shows
the distribution for $ | y(cm)/y_{Proj}(cm) | < 0.4$ and the full line
with circles shows the effect of an additional application of a cut in
transverse momentum  $0.2$GeV/c$\le p_T\le 0.8$GeV/c.
We see that each cut yields for slightly stronger peak of the squeeze. 
In the calculations of fig. \ref{squeeze-b} both cuts have been applied in order
to allow a better comparison to experimental data of the KaoS collaboration 
\cite{Foerster}. 

The right hand side of figure \ref{squeeze-p} compares the azimuthal distribution
at mid-rapidity (and with the given $p_T$-cut of $0.2$GeV/c$\le p_T\le 0.8$GeV/c) 
obtained in a calculation with (full line with circles) and without (dashed line
with squares) KN potentials, both using full re-scattering of the kaons. 
We see a visible increase of the peak when KN potentials
are active. If we use a calculation with KN-potentials 
but disable re-scattering (dotted line with triangles) we see a slight decrease
of the squeeze signal if compared to the calculation with full re-scattering
(full line with circles). 
However, it should be kept in mind, that the influence of the cuts (left hand side)
is of about the same size than the discussed effects.

If we now look for the sources, i.e. for the initial momenta of
the kaons of the production the curve is flat or even slightly convex.
A calculation without potential and re-scattering yields a similarly flat
distribution. We conclude that the influence of the nuclear medium - as well
potential as re-scattering - is important for explaining the squeeze, both
effects contribute.
Calculations of \cite{LiKo,Wang} see a stronger contribution of the potentials which
might perhaps be due to implementations of Lorentz-type forces.

\section{Conclusion}
We have analyzed azimuthal distributions and energy spectra of kaons in
Au+Au collisions at 1.48 AGeV. We find that the KN potential interactions 
only show a significant influence on the absolute yield of the kaons, which
is due to the penalty to be paid when producing a kaon in a dense medium.
Concerning the spectra only slight effects in the low momentum range can be
reported. However, the dynamics of the collision reaction affects the spectra.
Early reactions contribute stronger at high momenta while late reactions 
contribute more to the low momentum part. Thus, when decomposing the spectra 
into different time windows we find the spectra become steeper the later the 
time window is placed. 
Concerning the azimuthal distributions at mid-rapidity we see a visible influence
of the KN potential and of the re-scattering. The influence of re-scattering is
in coherence with the centrality dependence of this observable showing a 
maximum signal at peripheral collisions.
In conclusion the shape of spectra and azimuthal distributions give us
information about the collision dynamics rather than on a KN optical potential.


\section*{References}

\begin{thebibliography}{99}
\bibitem{Barth} R. Barth et al., (KaoS  Collaboration),
Phys. Rev. Lett. {\bf 78} (1997) 4007.
\bibitem{Menzel} M.~Menzel et al., (KaoS  Collaboration),
Phys. Lett. {\bf B 495} (2000) 26;\\
M. Menzel, Dissertation, Universit\"at Marburg, 2000.
\bibitem{Ritman} J.L. Ritman et al. (FOPI collaboration),  Z. Phys. {\bf A352} (1995) 355
\bibitem{Crochet}  P. Crochet et al. (FOPI collaboration), Phys. Lett. {\bf B486} 
(2000) 6
\bibitem{Devismes} A. Devismes et al. (FOPI collaboration),
J. Phys. G {\bf 28} (2002) 1591.
\bibitem{Laue} F. Laue 
et al., (KaoS  Collaboration), Phys. Rev. Lett. {\bf 82} (1999) 1640.
\bibitem{Sturm} C. Sturm et al (KaoS Collaboration), Phys. Rev. Lett. {\bf 86}
(2001) 39 \\
C. Sturm et al (KaoS Collaboration), J. Phys. G {\bf 28} (2002) 1895
\bibitem{Foerster} A. F\"orster  
et al., (KaoS  Collaboration), submitted to Phys. Rev. Lett. \\
 A. F\"orster  
et al., (KaoS  Collaboration), these proceedings
\bibitem{schaffi} J. Schaffner-Bielich et al., Nucl Phys {\bf A669} (2000) 153; 
\bibitem{Cassing}W.~Cassing et al., Nucl.~Phys.~A{\bf 614} (1997) 415. \\
E. Bratkovskaja et al., Nucl. Phys. {\bf A622} (1997) 593 \\
W. Cassing and E. Bratkovskaja, {\it Phys. Rep} {\bf 308} (1999) 65 
\bibitem{CLE00} J. Cleymans, H. Oeschler and K. Redlich,
Phys. Lett.~{\bf B485} (2000) 27.
\bibitem{ho_s2000} H.~Oeschler, J.~Phys.~G:  {\bf 27} (2001) 1.
\bibitem{Ko01} 
C.M. Ko, G.Q Li, J. Phys. G {\bf 22} (1996) 1673. \\
C.M. Ko, J.~Phys.~G {\bf 27} (2001) 327.
\bibitem{Fuchs} 
C. Fuchs et al. Phys. Lett {\bf B 434} (1998) 245, \\
C. Fuchs et al. Phys. Rev. Lett {\bf 86} (2001) 1794 \\
C. Fuchs et al. J. Phys. G {\bf 28} (2002) 1615 
\bibitem{LiKo} 
G Q Li et al.  Phys. Lett {\bf B 381} (1996) 17 
\bibitem{Wang} 
Z.S. Wang et al. Eur. Phys. J {\bf A 5} (1999) 275 
\bibitem{iqmd}C. Hartnack et al., Eur.Phys.J. {\bf A1} (1998) 151.
\bibitem{bass} S.A. Bass et al. Phys. Rev. {\bf C50} (1994) 2167.\\ 
S.A. Bass et al. Phys. Rev. {\bf C51} (1995) 3343.
\bibitem{juergen}J. Schaffner et al., Nucl. Phys. {\bf A625} (1997) 325.
\bibitem{ikaon} C. Hartnack and J. Aichelin, Proc. Int. 
Workshop XXVIII 
on Gross prop. of Nucl. and Nucl. Excit., Hirschegg, January 2000
edt by M. Buballa, W. N\"orenberg, B. Sch\"afer and J. Wambach \\
C. Hartnack and J. Aichelin, J. Phys. G {\bf 27} (2001) 571
\bibitem{kminus}
C. Hartnack, H. Oeschler and J. Aichelin,  Phys.Rev.Lett. {\bf 90} (2003) 102302.  
\bibitem{sqm2001}
C. Hartnack and J. Aichelin, J. Phys. G {\bf 28} (2002) 1649
\bibitem{oldhartnack} C. Hartnack et al., Nucl. Phys. {\bf A580} (1994) 643.

\end{thebibliography}
\end{document}